\documentclass[twocolumn,showpacs]{revtex4-1}
\usepackage{amsfonts}
\usepackage{amsmath}
\usepackage{graphicx}
\usepackage{graphics}

\begin{document}
\draft

\title{dc and ac magnetic properties of thin-walled Nb cylinders with and without a row of antidots}
\author{M.I. Tsindlekht$^1$, V.M. Genkin$^1$, I. Felner$^1$, F. Zeides$^1$, N. Katz$^1$, $\check{\text{S}}$.~Gazi$^2$, $\check{\text{S}}$. Chromik$^2$, O.V. Dobrovolskiy$^{3,4}$, R. Sachser$^3$, and M. Huth$^3$ }

\affiliation{$^1$The Racah Institute of Physics, The Hebrew University of Jerusalem, 91904 Jerusalem, Israel}

\affiliation{$^2$The Institute of Electrical Engineering SAS, D$\acute{u}$bravsk$\acute{a}$ cesta 9, 84104  Bratislava, Slovakia}

\affiliation{$^3$Physikalisches Institut, Goethe University, 60438 Frankfurt am Main, Germany}

\affiliation{$^4$Physics Department, V. Karazin Kharkiv National University, 61077 Kharkiv, Ukraine}

\begin{abstract}

dc and ac magnetic properties of two thin-walled superconducting Nb cylinders with a rectangular cross-section are reported. Magnetization curves and the ac response were studied on as-prepared and patterned samples in magnetic fields parallel to the cylinder axis. A row of micron-sized antidots (holes) was made in the film along the cylinder axis. Avalanche-like jumps of the magnetization are observed for both samples at low temperatures for magnetic fields not only above $H_{c1}$, but in fields lower than $H_{c1}$ in the vortex-free region. The positions of the jumps are not reproducible and they change from one experiment to another, resembling vortex lattice instabilities usually observed for magnetic fields larger than $H_{c1}$.
At temperatures above $0.66T_c$ and $0.78T_c$ the magnetization curves become smooth for the patterned and the as-prepared samples, respectively. The magnetization curve of a reference planar Nb film in the parallel field geometry does not exhibit jumps in the entire range of accessible temperatures.
The ac response was measured in constant and swept dc magnetic field modes. Experiment shows that ac losses at low magnetic fields in a swept field mode are smaller for the patterned sample. For both samples the shapes of the field dependences of losses and the amplitude of the third harmonic are the same in constant and swept field near $H_{c3}$. This similarity does not exist at low fields in a swept mode.

\end{abstract}
\pacs{74.25.F-, 74.25.Op, 74.70.Ad}
\date{\today}
\maketitle

\section{Introduction}

Penetration of magnetic flux into hollow superconducting cylinders is a long standing field of interest. The Little-Parks effect and the quantization of trapped flux were intensively studied during the last fifty years~\cite{LITTLE,DOU,VEKHT}. Recent advances in nanotechnology have made it possible for studying experimentally superconducting properties of thin films with different arrays of antidots, see for example,~\cite{Motta1} and references therein. In particular, for the observation of the aforementioned effects, cylinders or antidots of small diameter are required. At the same time, the study of hollow thin-walled cylinders with macroscopic sizes in magnetic fields parallel to its axis has been much less well studied. It was expected that quantum phenomena cannot be observed in such samples
because of the fact that one flux quanta for cylinders with a cross section area of $\approx 1$ cm$^2$ corresponds to a magnetic field about $10^{-7}$ Oe. In this case magnetization will be a smooth function of the magnetic field. However, experimental results obtained recently for thin-walled macroscopic cylinders do not agree with this expectation. Namely, in such Nb cylinders we succeeded in monitoring the magnetic moment of the current circulating in the walls and observed dc magnetic moment jumps even in fields much lower than $H_{c1}$ of the film itself~\cite{Katz1}.
So far it is not clear what mechanism is responsible for such flux jumps. Under an axial magnetic field the cylinder walls screen weak external fields, provided that $L\equiv Rd/\lambda^{2} \gg 1$, where $R$ is the cylinder radius, $d$ is the wall thickness, and $\lambda$ is the London penetration depth ~\cite{DOU,PG,KITTEL}. Therefore, it is expected, that a dc magnetic field, $H_0$, will penetrate into the cylinder as soon as the current in the wall exceeds the critical current and no field penetration should be observed at lower fields. Only above $H_{c1}$, vortices created at the outer cylinder surface can move into the cylinder. For a magnetic field oriented perpendicular to the Nb film surface such vortex motion leads to flux jumps~\cite{NOWAK,STAM}. These flux jumps were interpreted as a thermomagnetic instability of the critical state. It was demonstrated that in a sample with an array of antidots flux jump propagates along the antidots row \cite{MOTTA2}.

Nucleation of the superconducting phase in a thin surface sheath in decreasing magnetic fields parallel to the sample surface was predicted by Saint-James and de Gennes~\cite{DSJ}. They showed that nucleation occurs in a magnetic field $H_0\leq H_{c3}\approx 1.695 H_{c2}$. Experimental confirmations of this prediction were obtained soon after their work appeared. The experimental methods for this confirmation were dc resistivity and ac susceptibility measurements~\cite{ROLL}. It was found that low frequency losses in superconductors in surface superconducting states (SSS) can exceed losses in the normal state~\cite{ BURGER,ROLL}.

A swept dc magnetic field qualitatively changes the character of the ac response. Specifically, the penetration of the ac magnetic field into the sample takes place not only for $H_{c2}<H_0<H_{c3}$ but also for $H_{c1}<H_0<H_{c2}$, in sharp contrast to the case of constant dc fields~\cite{STR2,MAX,GENKIN22}. The effect of a swept dc field can more suitably be investigated by using hollow thin-walled superconducting cylinders, rather than by bulk samples, because one can control the field transmission through their walls. Previously, we have shown~\cite{Genkin1} that in a thin-walled cylinder in the mixed state, the effect of sweeping a dc field on the ac response is due to an enhancement of the vortex motion through the wall. Above $H_{c2}$, however, this picture is no longer appropriate and the experimental data were explained within the framework of a simple relaxation model~\cite{Katz1}.

The goal of this paper is to study how antidots affect the penetration of dc and ac magnetic fields into thin-walled superconducting Nb cylinders of macroscopic sizes, with a rectangular cross section.
 We show that at low enough temperatures for both, a flat and a patterned samples, even in the \emph{vortex-free regime} at $H< H_{c1}$, the dc magnetic field penetrates through the cylinder walls in an ``\emph{avalanche}''-like fashion. Jumps of the dc magnetic moment also become apparent at fields above $H_{c1}$ at low temperatures. For both samples, the field values at which jumps occur vary from one measurement to another, indicating that one deals with transitions between metastable states. At temperatures above $0.66T_c$ and $0.78T_c$ the magnetization curves become smooth for the patterned and the as-prepared sample, respectively.

The ac response of both cylinders was studied in the point-by-point and swept field modes. In these, the signals of the first, second and third harmonics were measured concurrently. The ac response of as-prepared and patterned samples is qualitatively different in a swept field mode.

\section{Experimental}

\begin{figure}
\begin{center}
\leavevmode
\includegraphics[width=0.6\linewidth]{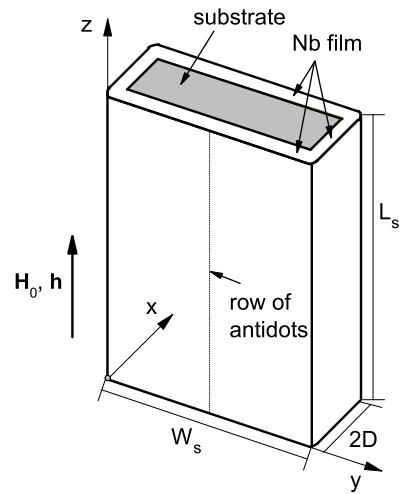}
\caption{Sketch of the $B$ sample. Here $\text{L}_s = 7.5$ mm, $\text{W}_s=3$ mm, and $2\text{D}=1.4$ mm are the substrate length, width and thickness, respectively. Both dc and ac fields were parallel to $Z$-axis. Dimensions are not to scale.}
\label{f1}
\end{center}
\end{figure}

The cylindrical samples were prepared by dc magnetron sputtering at room temperature on a rotated sapphire substrate. The sizes of the substrate with rounded corners (radius 0.2 mm) are $1.5\times3\times7.5$ mm$^3$. We fabricated, therefore, a thin-walled hollow superconducting cylinder with a rectangular cross section. The nominal film thickness of both samples was $d=100$ nm. A sketch of the sample geometry is presented in Fig.~\ref{f1}.

\begin{figure}
\centering
       \includegraphics[width=0.94\linewidth]{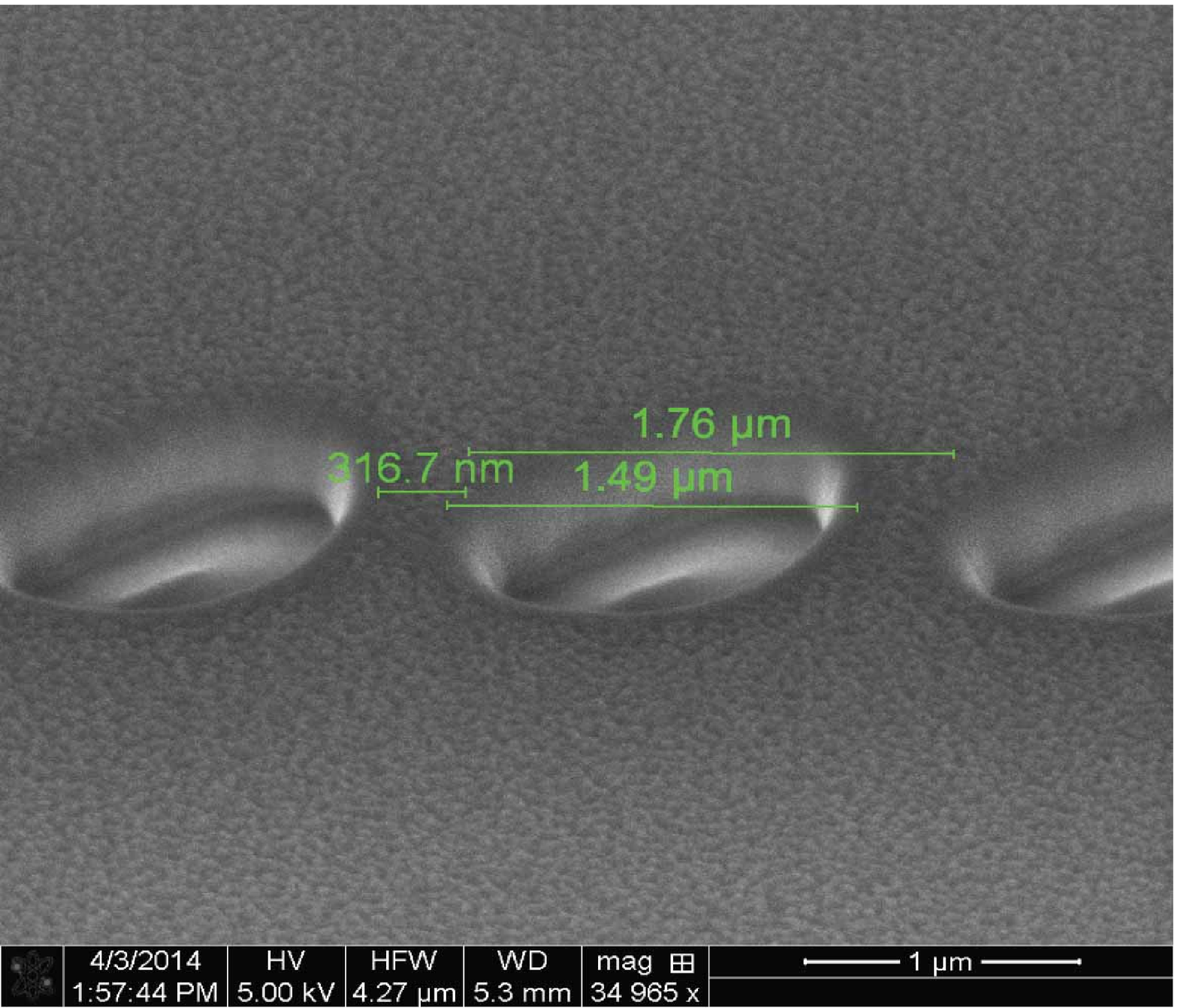}
      \includegraphics[width=0.94\linewidth]{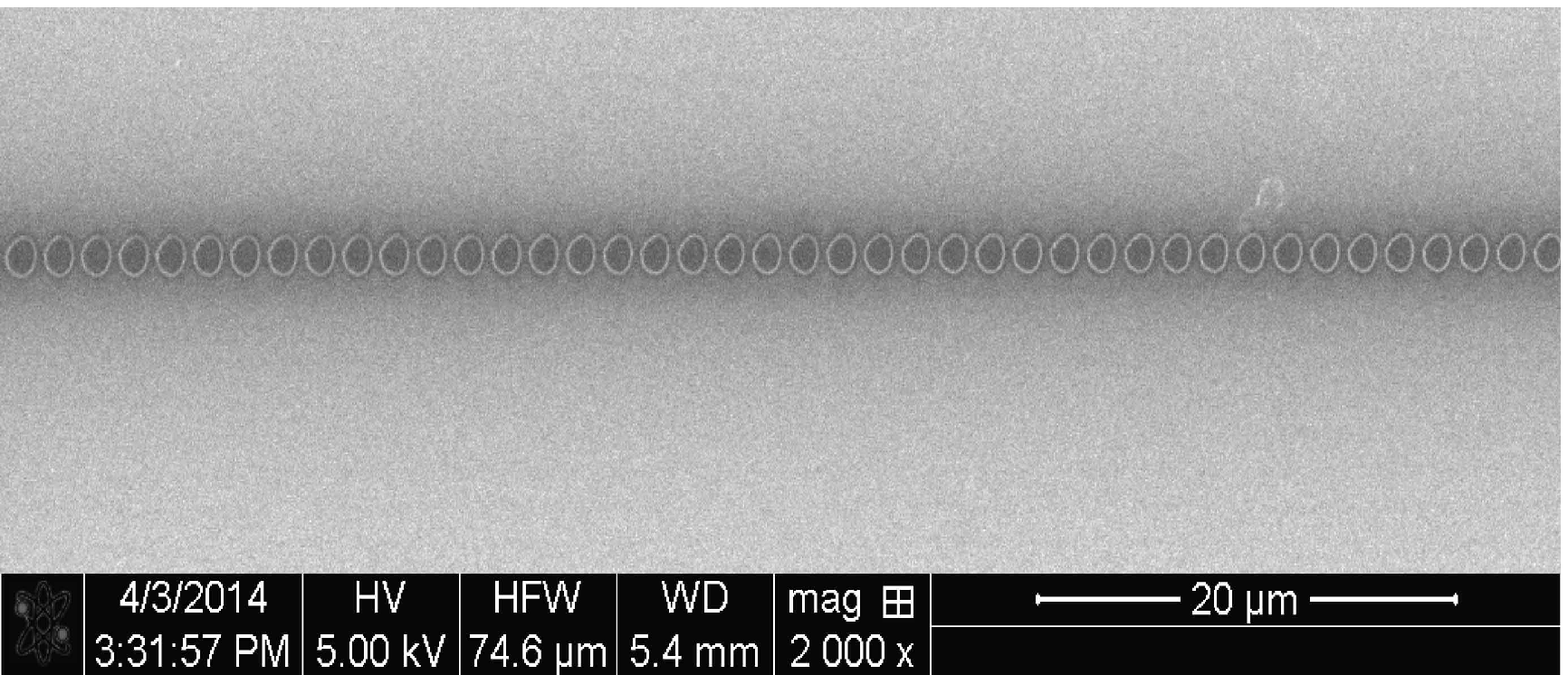}
    \caption{SEM images of the surface of sample $B$. The antidots have an average diameter of 1.5 $\mu$m and an average edge-to-edge distance of 300 nm. An overview SEM image is presented in the bottom panel where the row of FIB-milled antidots is clearly seen. }
    \label{f2}
\end{figure}

The reference sample $A$ was kept as-grown, while the second one, sample $B$, was patterned with a row of antidots at the mid of the larger surface over the entire length of the sample.
The row of antidots was milled by focused ion beam (FIB) in a scanning electron microscope (FEI, Nova Nanolab 600). The beam parameters were 30\,kV/0.5\,nA, while the defocus and blur were 560\,$\mu$m and 3\,$\mu$m, respectively. The pitch was equal to the antidot center-to-center distance of 1.8\,$\mu$m and the number of beam passes needed to mill 150\,nm-deep antidots was 2000. The antidots row with a length of $7.5$\,mm was milled by iteratively stitching the processing window with a long size of $400\,\mu$m. SEM images of the patterned surface of sample $B$ are shown in Fig.~\ref{f2}. The antidots have an average diameter of 1.5\,$\mu$m and an average edge-to-edge distance of 300\,nm.

The dc magnetic properties were measured using a commercial superconducting quantum interference device (SQUID), Quantum Desing MPMS5, magnetometer. The ac response was measured by the pick-up coil method. The sample was inserted into one coil of a balanced pair of coils, and the unbalanced signal was measured by means of lock-in amplifier. The ac magnetic susceptibilities were measured in absolute units, see~\cite{LEV2}. A ``home-made'' measurement cell of the experimental setup was adapted to the SQUID magnetometer. A block diagram of the experimental setup can be found elsewhere~\cite{LEV2}.

The ac response as a function of the dc field were carried out by two methods: (i) - point-by-point (PBP) mode, where the dc field was kept constant during the measurement, and (ii) - swept field (SF) mode, where the dc field was ramped with a rate of 20 Oe/s. Both external ac and dc fields were directed parallel to cylinder axis and hence, to the film surface.

\section{Results}

\subsection{dc magnetization}

\begin{figure}
    \centering
    \includegraphics[width=0.98\linewidth]{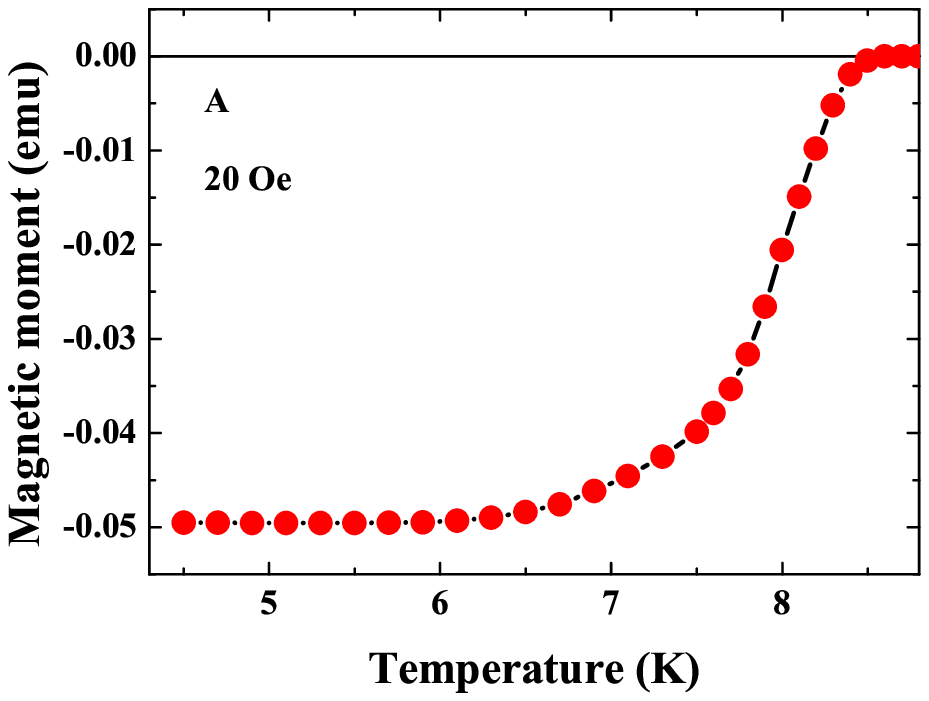}
    \includegraphics[width=0.98\linewidth]{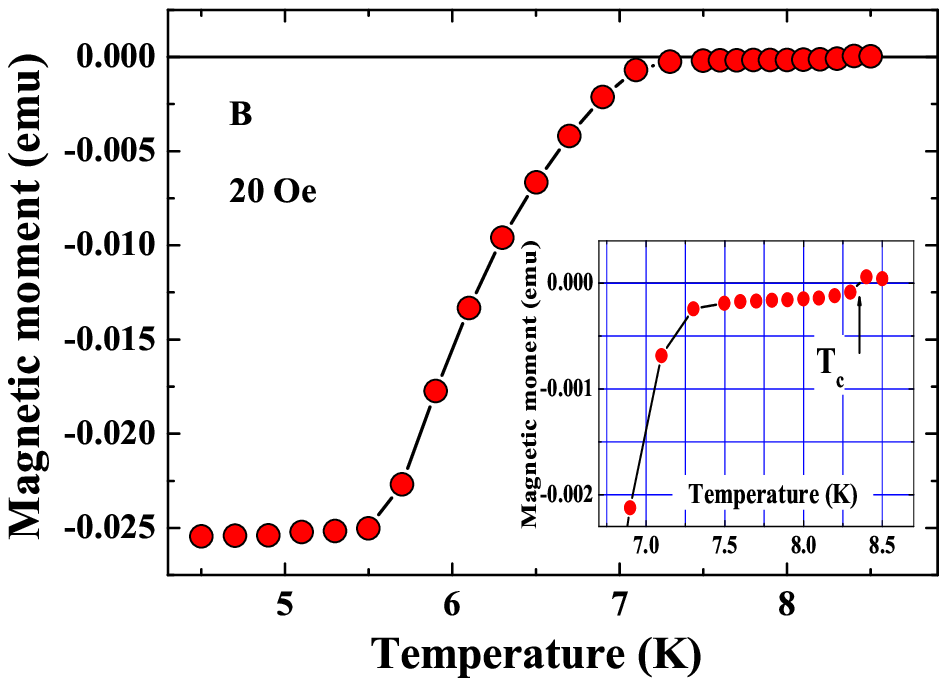}
    \caption{(Color online). Temperature dependences of the magnetic moment of samples $A$ and $B$, upper and lower panels, respectively. Inset to lower panel shows temperature dependence of $M_0$ of $B$ sample near $T_c$. }
    \label{f3}
\end{figure}

The upper and lower panels of Fig.~\ref{f3} show the temperature dependences of the magnetic moments, $M_0$, in magnetic field $20\pm 2$ Oe, of samples $A$ and $B$, respectively. The critical temperatures, $T_c$, of both samples are almost the same, 8.3 K, the transition width for sample $A$ is 1.3 K but 2.7 K for sample $B$. Sample $B$ demonstrates a two-stage transition, see the inset to the lower panel of Fig.~\ref{f3}.
At low temperatures, the magnetic moment of sample $A$ is a factor of two larger than that of sample $B$.
Temperature and field dependences of the magnetic moment were measured after cooling the sample down to the desired temperatures in zero field (ZFC).

\begin{figure}
    \centering
    \includegraphics[width=0.98\linewidth]{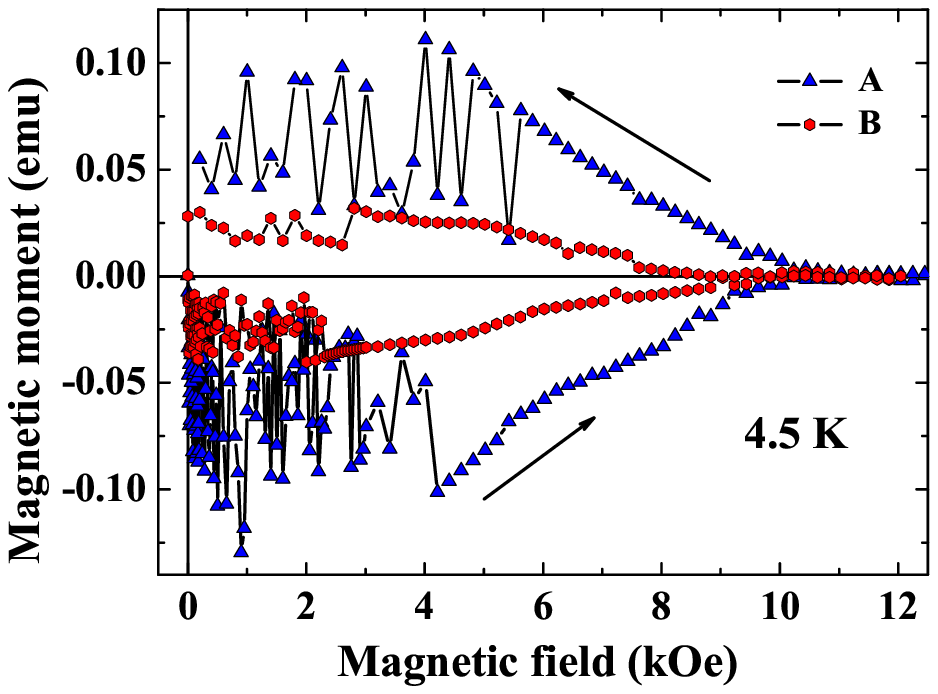}
    \includegraphics[width=0.98\linewidth]{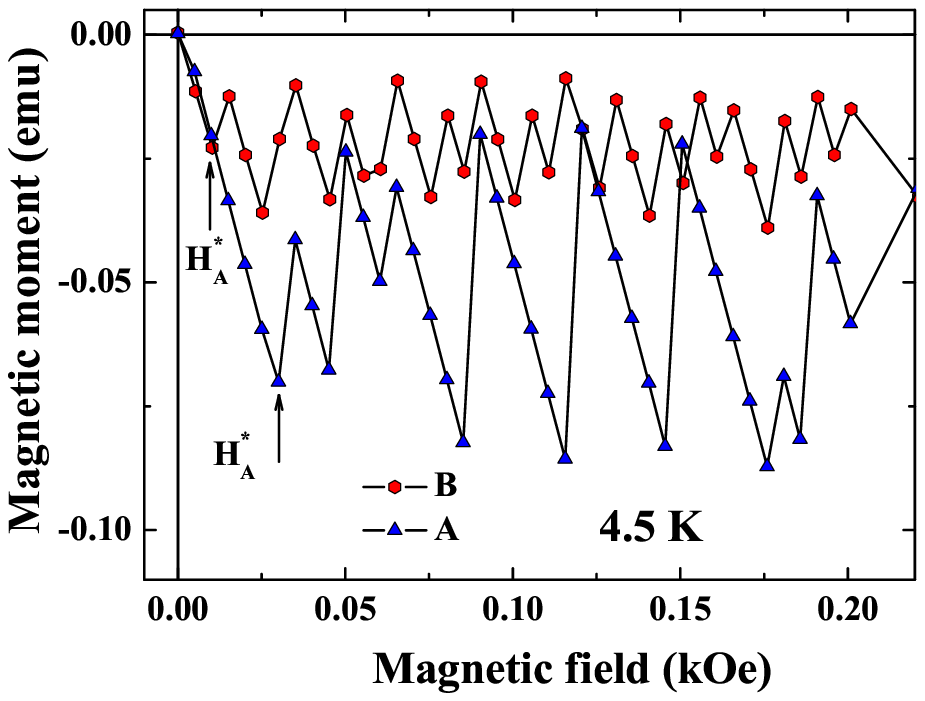}
    \caption{(Color online). $M_0(H_0)$ of samples $A$ and $B$ after ZFC, upper panel. Expanded view of the magnetization curves in low magnetic fields for samples $A$ and $B$, lower panel. }
    \label{f5}
\end{figure}

The $M_0(H_0)$ dependences for samples $A$ and $B$ at 4.5 K are shown in the upper panel of Fig.~\ref{f5}. The magnetization curves in the ascending branch were measured in the hysteresis mode with 5 Oe step at low fields. Fig.~\ref{f5} shows that the $H_{c2}$ values are different. Determination of $H_{c2}$ for sample $B$ is less accurate than that of sample $A$, due to the magnetic moment relaxation, which at high fields is larger for sample $B$~\cite{MIT}.
 An expanded view of the magnetization curves at low fields is shown in the lower panel of Fig.~\ref{f5}. The fields of the first jumps, $H^*$, are around 20 Oe and 10 Oe, while the number of jumps in magnetic fields up to 100 Oe are 5 and 7 for samples $A$ and $B$, respectively. Jumps of the magnetic moment were observed in a wide range of magnetic fields, including fields below $H_{c1}$ for both samples. This behavior is reminiscent of magnetic flux jumps in Nb thin films for $H_0$ perpendicular to the film surface~\cite{NOWAK,STAM}. The jumps observed in these papers were interpreted as a thermomagnetic instability of the Abrikosov vortex lattice~\cite{NOWAK,STAM}. However, existence of jumps in fields below than $H_{c1}$ and parallel to the surface have been reported in our recent work only \cite{Katz1}. $H_{c1}$ is $\approx 350$ Oe at 4.5 K in our samples. Direct determination of $H_{c1}$ for thin-walled cylindrical samples is impossible due to magnetic moment jumps at low fields. However, the estimation of $H_{c1}$ can be done using magnetization curves of the planar film as it shown in inset to Fig.\ref{f13}.

\subsection{ac response}

The effective ac magnetic susceptibility of the sample in the external field $H(t)=H_0(t)+h_{ac}\sin (\omega t)$
is given by
  \begin{equation}\label{Eq2}
M(t)=Vh_{ac}\sum_n\{\chi_n^{'} \sin (n\omega t)-\chi_n^{''}\cos (n\omega t)\},
\end{equation}
and it exhibits the appearance of the ac field penetration into the sample, i.e. $\chi_1^{'}\neq -1/4\pi$, ac losses $\chi_1^{''}>0$  and harmonics of the fundamental frequency, $\chi_n$. Here, $M(t)$ is the magnetic moment of the sample and $V$ is its volume. In what follows we consider the results of the ac measurements in both PBP and SF modes.

\begin{figure}
\begin{center}
    \includegraphics[width=0.98\linewidth]{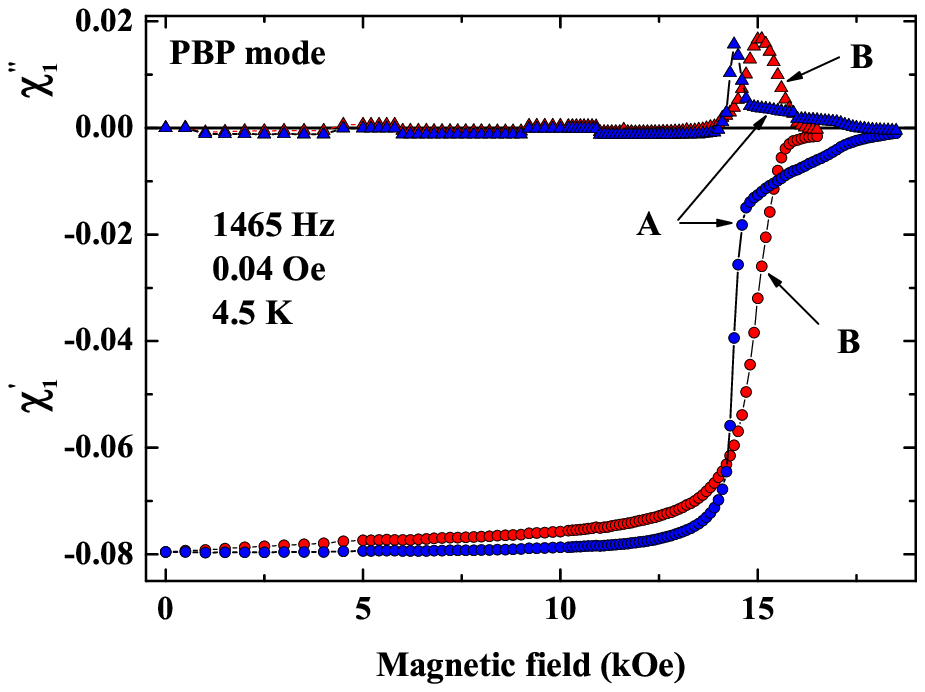}
    \includegraphics[width=0.98\linewidth]{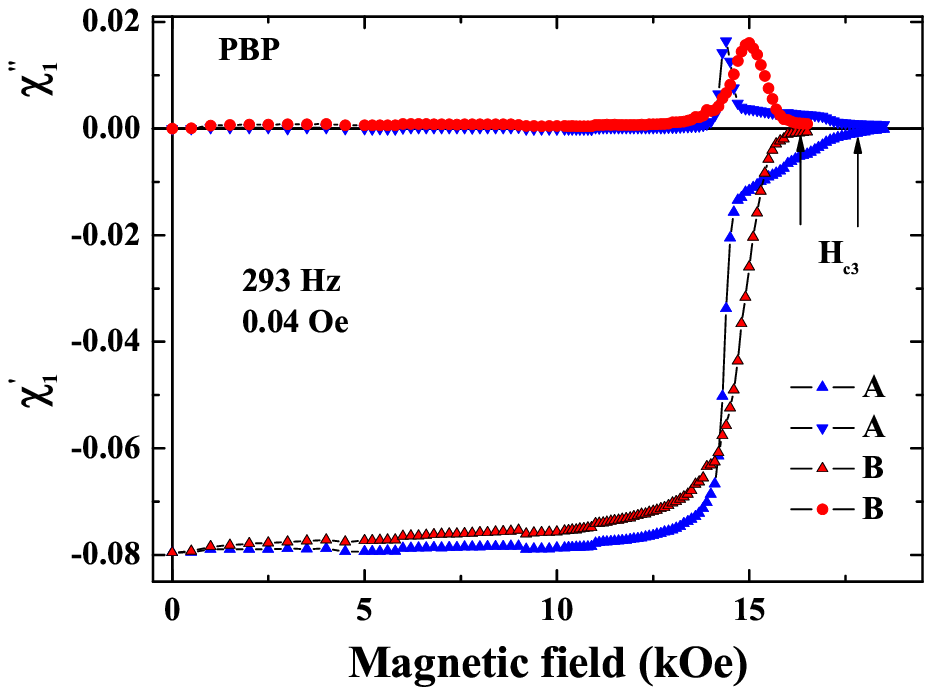}
    \caption{(Color online). Field dependences of $\chi_1(H_0)$ of samples $A$ and $B$ in the PBP mode at 1465 and 293 Hz, upper and lower panels, respectively. Measurements were done at 4.5 K. Arrows on the lower panel show $H_{c3}$ for both samples.}
    \label{f5aa}
\end{center}
\end{figure}

\begin{figure}
\begin{center}
\leavevmode
\includegraphics[width=0.9\linewidth]{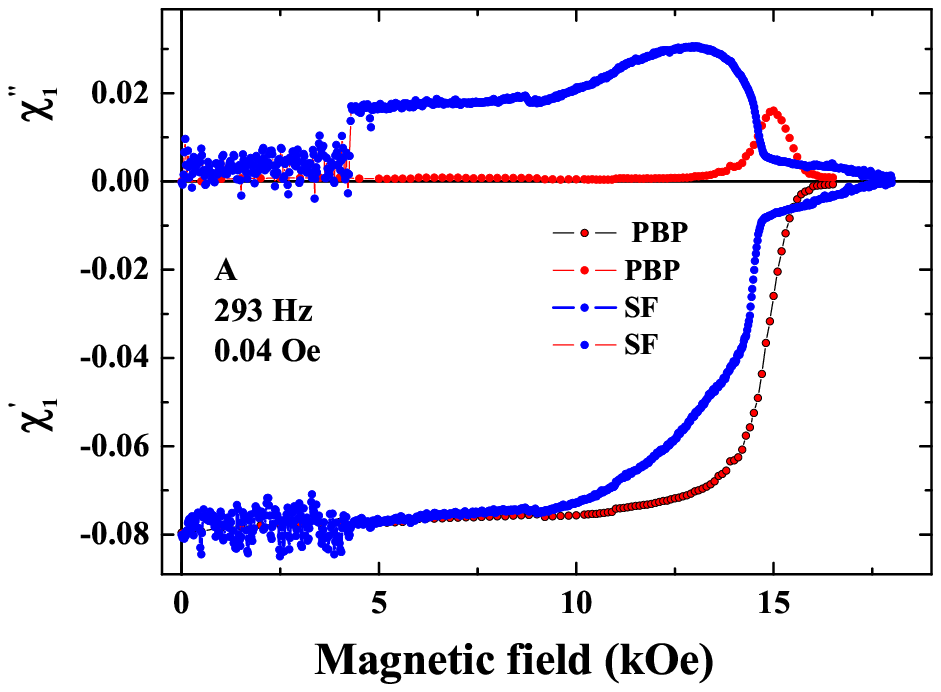}
\includegraphics[width=0.9\linewidth]{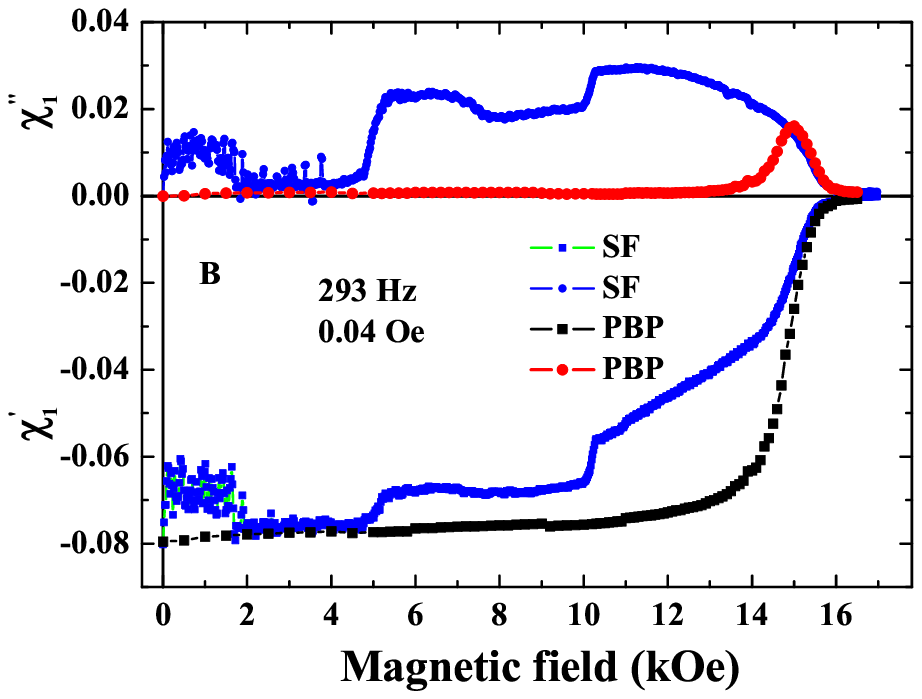}

\caption{(Color online). Field dependences of $\chi_3$ of samples $A$ and $B$  (upper and lower panels, respectively) in the PBP mode at 4.5 K.}
\label{f6a}
\end{center}
\end{figure}

The real and imaginary components of the ac susceptibility at 4.5 K for both samples measured in the PBP mode as a function of $H_0$ at two frequencies are shown in Fig.~\ref{f5aa}. Almost complete screening up to 12.5 kOe of the ac field by the superconducting walls is observed for both samples. This value is higher than $H_{c2} =11\pm 0.5$ kOe of sample $A$ (Fig.~\ref{f5}, upper panel). Complete screening of ac fields by a type II superconductor at low frequencies ($\omega \ll \omega_p$, here $\omega_p$ is a depinning frequency) and amplitudes of excitation (ac current much lower than depinning current) in dc fields lower than $H_{c2}$ was observed years ago \cite{STR2}.
The frequency dispersion of $\chi_1$ is weak for both samples. Third critical magnetic field was determined using ac data as follows. At low amplitude of excitation a loss peak located between $H_{c2}$ and $H_{c3}$. Losses disappear at $H_0>H_{c3}$ because in a normal state $\delta>>d$. Here $\delta$ is a skin depth in a normal state. Such determination of $H_{c3}$ was proposed years ago by Rollins and Silcox~\cite{ROLL}.
The lower panel of Fig.\ref{f5aa} shows an example of determination of the third critical magnetic field. It was found that $H_{c3}\approx 17.5\pm 0.5$ and $16\pm 0.5$ kOe at 4.5 K for $A$ and $B$ samples, respectively. $H_{c3}/H_{c2} \approx 1.6$ for sample $A$. An accurate determination of $H_{c2}$ for sample $B$ is difficult, due to magnetic relaxation, as discussed above. The absorption line, $\chi_1^{''}(H_0)$, near $H_{c3}$ is different for samples $A$ and $B$. Thus, this line is nonuniform for sample $A$ and it is uniform but broadened for sample $B$.
The ac response of superconductors even at very low amplitude of excitation, e.g., less than 1 Oe, is strongly nonlinear in the SSS \cite{ROLL,Genkin1}. The second harmonic signal is absent in the PBP mode in the entire range of magnetic fields. At the same time, the third-harmonic signal exists in the vicinity of $H_{c3}$ only. The absence of the second harmonic in PBP mode is a common feature for the bulk samples as well~\cite{CAMP}. Fig. \ref{f6a} shows the field dependences of $\chi_3$, $\chi_{2,3}\equiv \sqrt{(\chi_{2,3}^{'})^{2}+(\chi_{2,3}^{''})^{2}}$, in PBP mode
for samples $A$ and $B$, in the upper and lower panels, respectively. Perturbation theory with respect to the amplitude of excitation is not applicable for interpreting these experimental data. For example, according to perturbation theory, $\chi_3$ should be proportional $h_{ac}^{2}$ and this is not the case in our findings, Fig. \ref{f6a}. It is known that perturbation theory cannot explain experimental data for bulk samples too \cite{ROLL,GENKIN22}. We also note that there is a difference for the third harmonic signal between samples $A$ and $B$ in the PBP mode.

\begin{figure}
 \begin{center}
    \includegraphics[width=0.98\linewidth]{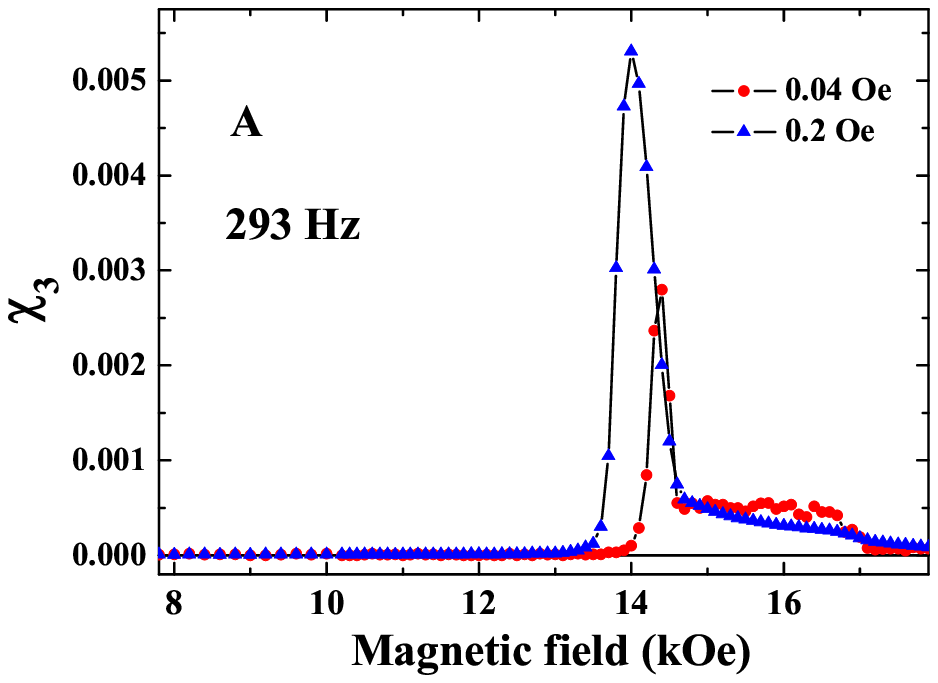}
    \includegraphics[width=0.98\linewidth]{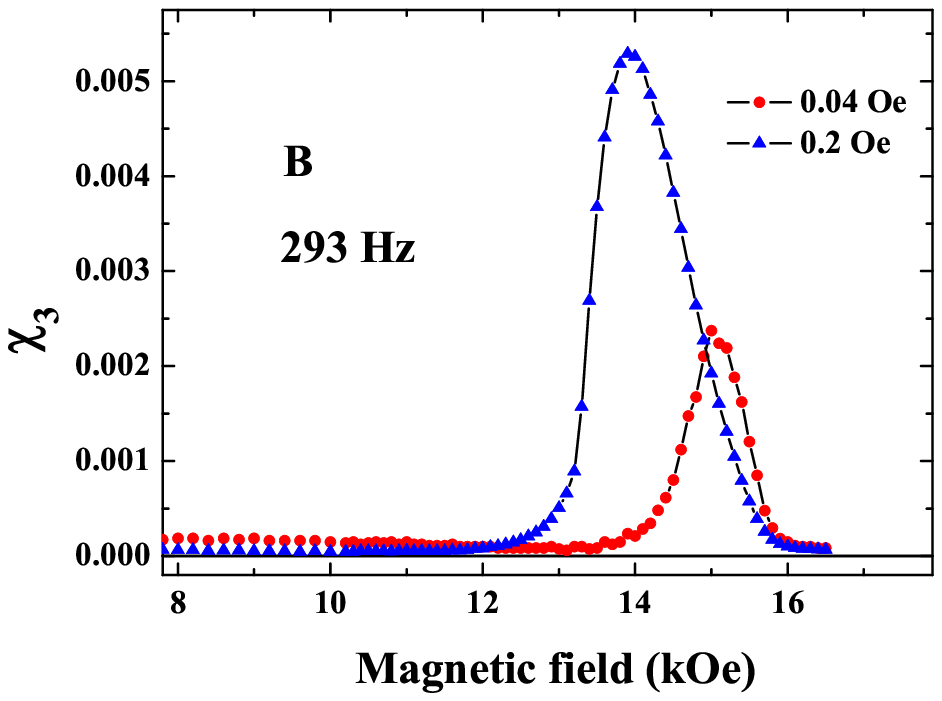}
    \caption{(Color online). Field dependences of $\chi_1$ for samples $A$ and $B$ (upper and lower panels, respectively) in the PBP and SF modes at 293 Hz and an excitation amplitude of 0.04 Oe. Measurements were done at 4.5 K.}
    \label{f7c}
\end{center}
\end{figure}

A swept field affects the ac response more strongly at low frequencies or/and low excitation amplitudes for a given sweep rate. This was confirmed in experiments with bulk and thin-walled cylinders samples \cite{MAX,GENKIN22} and \cite{Genkin1,Katz1}, respectively. Fig. \ref{f7c} shows the field dependences $\chi_1$ for both samples $A$ and $B$ in the PBP and SF modes at 293 Hz and amplitude 0.04 Oe. The difference between the PBP and SF modes can easily be seen for both samples. The ac response at low magnetic fields in the SF mode are fluctuating due to magnetic flux jumps, Fig.~\ref{f5}. Near $H_{c3}$ the curves of $\chi_1$ coincide well in PBP and SF modes for both samples, Fig. \ref{f7c}. The difference between the two samples in the SF mode is very pronounced in fields above 5 kOe. In particular, $\chi_1^{''}$ is a smooth function of the dc field for sample $A$, but for sample $B$ it shows step-like features in fields near 7 and 10 kOe.

\begin{figure}
\begin{center}
\leavevmode
\includegraphics[width=0.9\linewidth]{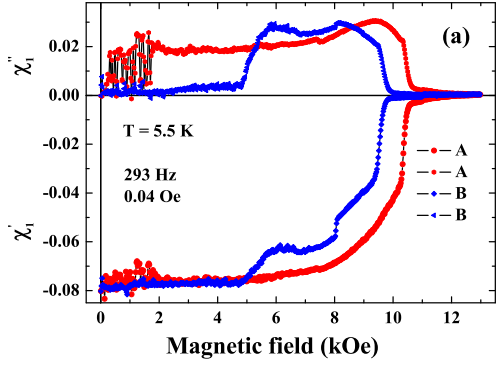}
\includegraphics[width=0.9\linewidth]{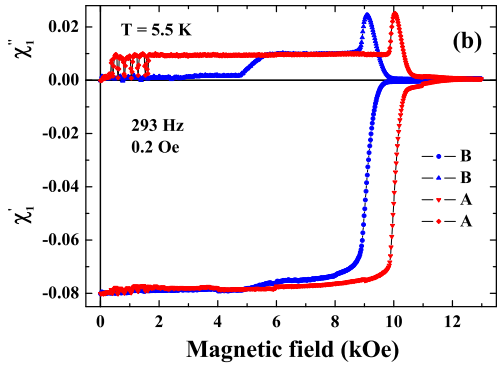}
\includegraphics[width=0.9\linewidth]{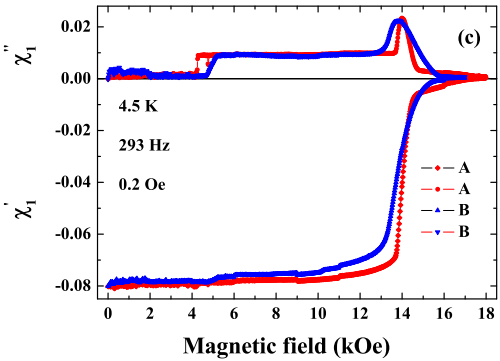}
\caption{(Color online). Field dependences of $\chi_1$ for samples $A$ and $B$  in the SF mode at 5.5 K (panels $a$ and $b$, respectively) and 4.5 K (panel $c$).}
\label{f11a}
\end{center}
\end{figure}

\begin{figure}
\begin{center}
\leavevmode
\includegraphics[width=0.9\linewidth]{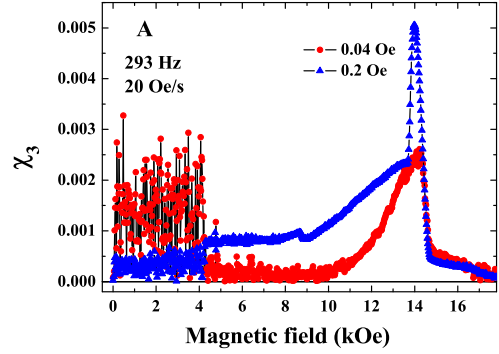}
\includegraphics[width=0.9\linewidth]{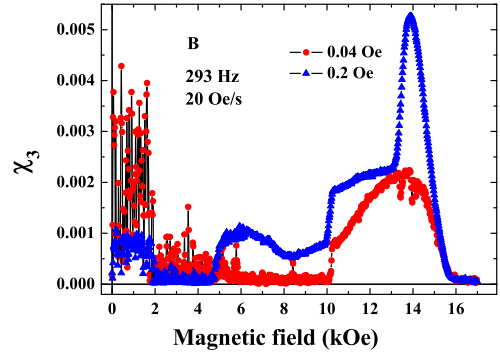}

\caption{(Color online). Field dependences of $\chi_3$ of $A$ and $B$ samples (upper and lower panels, respectively) in the SF mode at 4.5 K.}
\label{f9a}
\end{center}
\end{figure}

\begin{figure}
\begin{center}
\leavevmode
\includegraphics[width=0.9\linewidth]{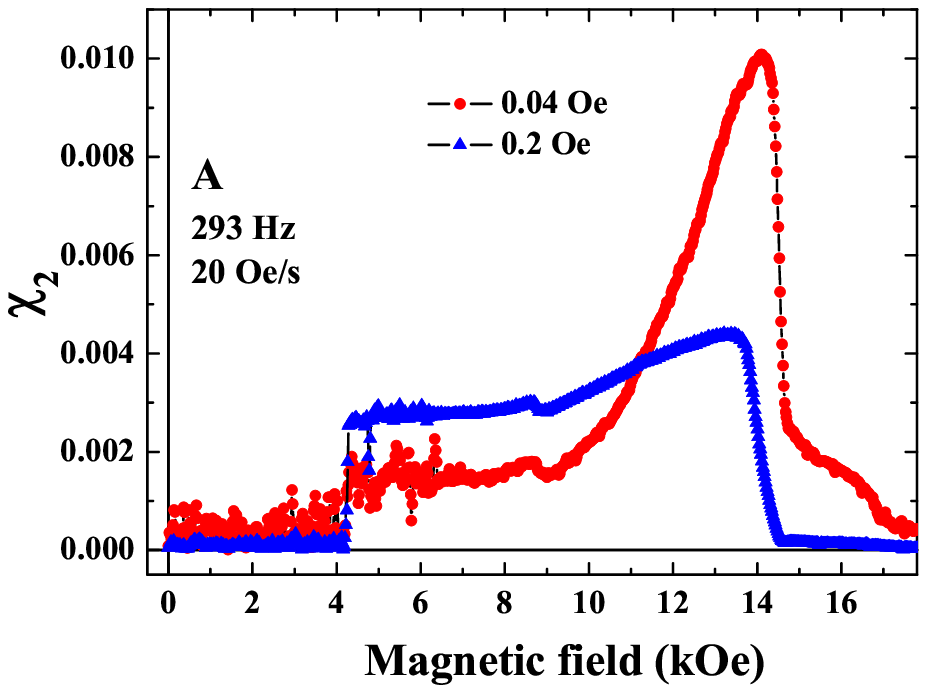}
\includegraphics[width=0.9\linewidth]{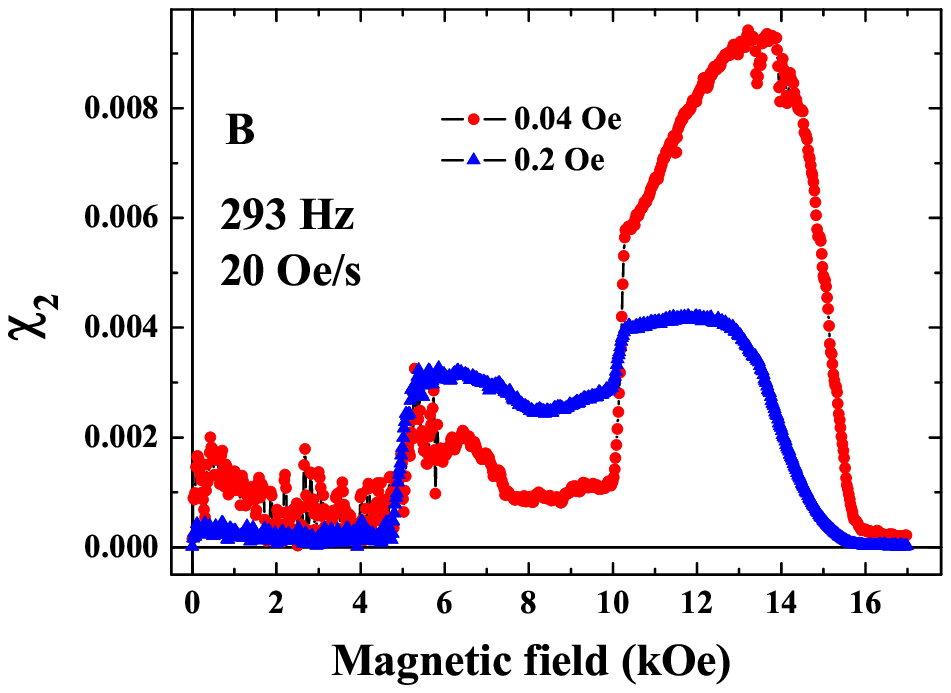}

\caption{(Color online). Field dependences of $\chi_2$ of $A$ and $B$ samples, upper and lower panels, respectively, in the SF mode at 4.5 K.}
\label{f10a}
\end{center}
\end{figure}

A nonlinearity can clearly be seen not only in the second and third harmonics, but in the first harmonic too. Fig. \ref{f11a} shows the field dependences of $\chi_1$ of samples $A$ and $B$ at $h_{ac}=0.04$ and 0.2 Oe and T = 4.5 and 5.5 K in the SF mode. Panels $a$ and $b$ demonstrate: (i) that at low magnetic fields, losses in sample $A$ are significantly larger than losses in sample $B$ and: (ii) an increase of the excitation amplitude leads to a decrease of $\chi_1{''}$ for both samples. At $h_{ac}= 0.2$ Oe for $H_0> 5$ kOe there is a plateau and $\chi_1^{''}$ for both samples coincides with high precision. The plateau in the SF mode at high excitation amplitudes was observed at T = 4.5 K, Fig. \ref{f11a}$c$ and also at 5.5 K for, Fig. \ref{f11a}$b$. It appears that in this range of magnetic fields and at high enough amplitude, the first harmonic signal of the two samples is almost identical. However, a qualitative difference remains for the signals of the second and third harmonics, see Figs.~\ref{f9a} and \ref{f10a}.

As for the second harmonic signal it is absent for both samples in the whole range of magnetic fields in the PBP mode, but becomes visible in the SF mode. Fig. \ref{f10a} shows the field dependences of $\chi_2$ in the SF mode. Perturbation theory cannot explain the data for $\chi_2$ in the SF mode and $\chi_3$ in both modes. In accordance to this theory one could expect that $\chi_3\propto h_{ac}^2$ and  $\chi_2\propto h_{ac}$. However, this is not the case in our experiment at any magnetic field. In our experiment, an increase of the excitation amplitude leads to a suppression of $\chi_2$. In the SF mode $\chi_2$ is larger than $\chi_3$ under the conditions of the experiment, see Figs. \ref{f9a} and \ref{f10a}. We note that the data for $\chi_1$, $\chi_2$ and $\chi_3$ fluctuate strongly at fields lower than 4 kOe at 4.5 K for sample $A$ due to magnetic flux jumps.

\begin{figure}
\begin{center}
\leavevmode
\includegraphics[width=0.9\linewidth]{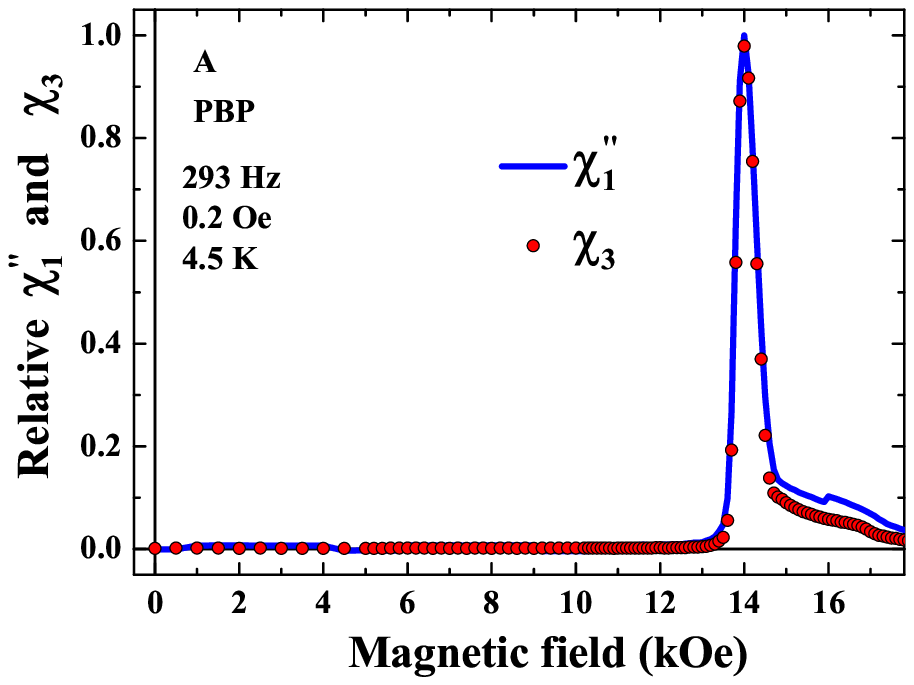}
\includegraphics[width=0.9\linewidth]{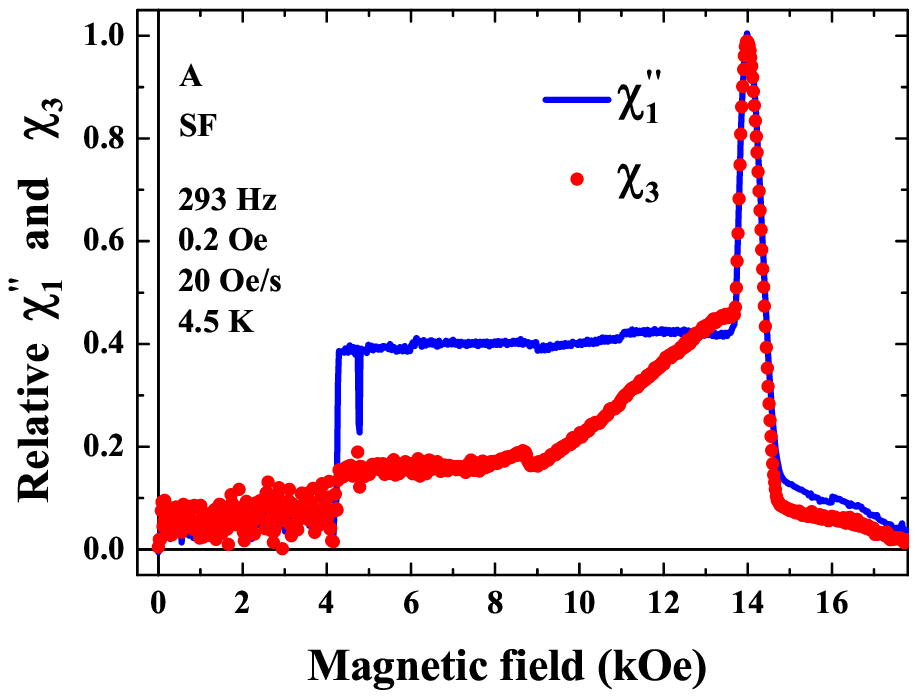}

\caption{(Color online). Field dependences of normalized $\chi_1^{''}$ and $\chi_3$ of sample $A$ in point-by-point and swept field modes (upper and lower panels, respectively). The shapes of $\chi_1^{"}$ and $\chi_3$ are with high accuracy identical in PBP and SF modes near $H_{c3}$. This similarity breaks in a SF mode at low magnetic fields.}
\label{f8c}
\end{center}
\end{figure}

\begin{figure}
\begin{center}
\leavevmode
\includegraphics[width=0.9\linewidth]{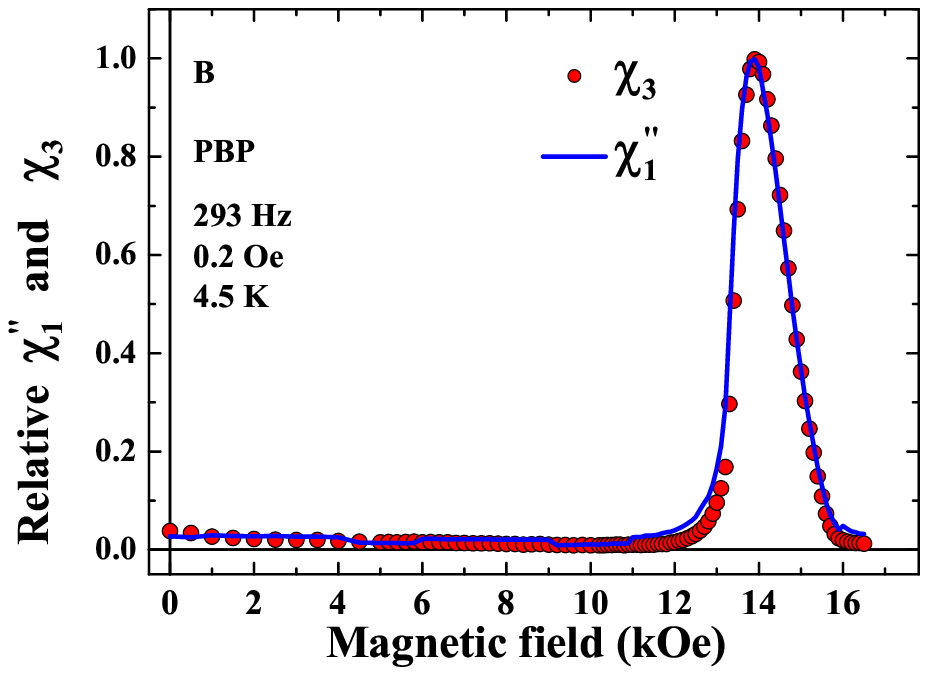}
\includegraphics[width=0.9\linewidth]{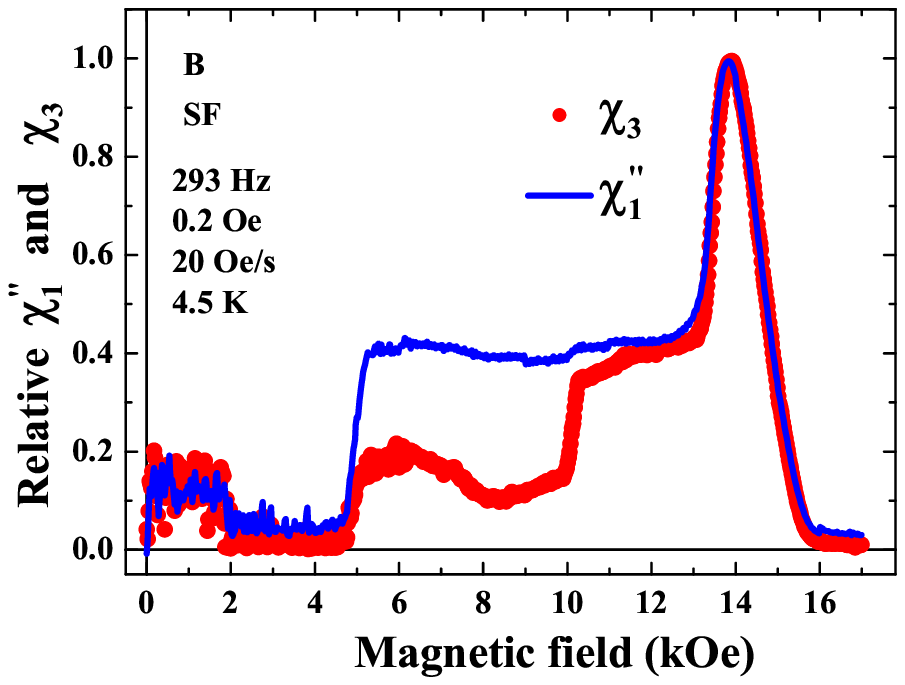}

\caption{(Color online). Field dependences of normalized of $\chi_1^{"}$ and $\chi_3$ of sample $B$ in point-by-point and swept field modes (upper and lower panels, respectively). The shapes of $\chi_1^{"}$ and $\chi_3$ are with high accuracy identical in PBP mode and in large fields in SF mode.}
\label{f8b}
\end{center}
\end{figure}

It is interesting to note the following concerning the relation between field dependences of $\chi_1^{''}$ and $\chi_3$. Figs.~\ref{f8c} and \ref{f8b} show field dependences of normalized $\chi_1^{''}$ and $\chi_3$ for samples $A$ and $B$. Upper panels in both figures correspond to the PBP mode and lower panels to SF mode. At low magnetic fields  $\chi_1^{''}$ and $\chi_3$ are very small in the PBP mode for both samples. Both signals become measurable near $H_{c3}$ and the shape of these signals is identical with high precision.  In the SF mode the shapes of  $\chi_1^{''}$ and $\chi_3$ are again the same in the vicinity of $H_{c3}$. However, at low magnetic fields this similarity vanishes in the SF mode. Such similarity in the PBP mode can be proved in the frame of perturbation theory~\cite{PAV}, but it has not yet proven in the general case which we face in our experiment.

\section{Discussion}

\subsection{dc magnetization curves}
\begin{figure}
\begin{center}
\leavevmode
\includegraphics[width=0.9\linewidth]{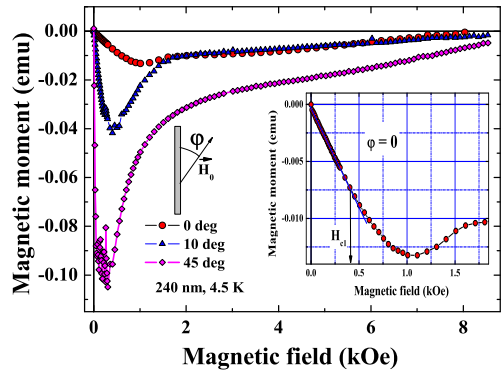}
\caption{Ascending branches of magnetization curves of planar film in parallel and tilted magnetic fields. Inset shows determination of $H_{c1}$ of the planar film.}
\label{f13}
\end{center}
\end{figure}

The physical reasons for the observed flux jumps at small magnetic fields are not clear. One can suggest that the alignment of the magnetic field with respect to the sample surface is not perfect. Indeed, the latter cannot be ruled out completely, and a small field component perpendicular to the surface, $H_{\bot}$, should create vortices which might be responsible for the flux jumps at small magnetic fields. Hence, one may expect that flux jumps could be present at small magnetic fields in a reference planar film as well. This assumption has been examined in an additional control experiment with a reference planar film. Figure \ref{f13} displays ascending branches of the magnetization curves of the planar Nb film of 240\,nm thickness sputtered onto a silicon substrate, for the magnetic field inclination angles $\varphi =0^{\circ}$, $10^{\circ}$, and $45^{\circ}$. For $\varphi = 10^{\circ}$ and $45^{\circ}$ the component $H_{\bot}\approx0.17H_0$ and $H_{\bot}\approx0.71H_0$, respectively. Vortices created by this field component exist at small magnetic fields. This experiment demonstrates that in small fields the magnetic moment is a linear function of the magnetic field value and vortices created by $H_{\bot}$ \emph{do not induce any flux jumps} at small fields. The magnetic moment at small fields remains a linear function of the magnetic field for planar films of different thicknesses. Magnetic moment jumps first appear in the magnetization curve at inclination angles larger than $10^\circ$. Such a field inclination angle is at least a factor of 3 larger than the orientational misalignment of the sample orientation with respect to the field direction in our experiment. Therefore, the results obtained for planar films suggest that the vortices created by the small field component perpendicular to the surface are not the cause for magnetic moment jumps at small magnetic fields in the cylindrical samples.

The experimental data demonstrate the existence of magnetic instabilities in fields lower than $H_{c1}$. At 4.5 K, the flux starts to penetrate into the cylinders $A$ and $B$ at $H_0 = 20$ and 10 Oe, respectively, Fig.~\ref{f5} (lower panel). The field of the first jump, $H^*$, is defined by the some critical current (not to be confused with a depairing current). If we assume that the critical current density in the isthmus between two antidots is the same as in the film, then the ratio  $H_B^*/ H_A^*$ should be $\approx 0.16$. However, the experiment shows that this ratio is about 0.5, see Fig.~\ref{f5}. This means that the critical current density in the isthmuses is higher than in the as-prepared film. We note that the ratio of the magnetic moments in ZFC in field 20 Oe  for samples $B$ and $A$ is 0.5, see Fig. \ref{f3}. In accordance to the thermodynamic criterion \cite{Katz1} $H^* \propto\sqrt{d}$. Comparison $H^*$ for $A$ sample and samples from \cite{Katz1} shows that the thermodynamics cannot describe these magnetization jumps in samples without antidots.

It was demonstrated that at low temperature and at magnetic fields higher than some critical value, $H_{th}$, the magnetization curve becomes smooth and $H_{th}$ is sufficiently larger in the sample with an array of antidots~\cite{Motta1}. The latter experiments were carried out with the field perpendicular to the film surface. In our case we deal with the row of antidots and the magnetic field parallel to the surface. We believe that this is the main reason why $H_{th}$ is lower for the sample with antidots, see upper panel of Fig. \ref{f5}. We have to mention that the difference between perpendicular and parallel geometries is crucial. For example vortex velocity in the perpendicular geometry is a few orders magnitude larger than for the parallel one, see Ref. \cite{Genkin1}.

\subsection{ac response}
The field dependences of $\chi_1(H_0)$ in the PBP mode are different for $A$ and $B$ samples, Fig. \ref{f5aa}. Losses appear and screening decreases in magnetic fields above $H_{c2}$. Near $H_{c3}$ there is a loss peak and the shape of this peak is different for samples $A$ and $B$. The shape of the loss peak for sample $A$ is nonuniform and for sample $B$ it is broadened. The third critical field of sample $A$ is larger than for sample $B$, Fig. \ref{f5aa} lower panel. However, the determination of $H_{c3}$ for sample $B$ is questionable.

The difference in the ac response of samples $A$ and $B$ becomes qualitative in the SF mode, Figs.~\ref{f7c}, \ref{f9a} and \ref{f10a}. Whereas the field dependences of $\chi_1$, $\chi_2$ and $\chi_3$ are smooth for sample $A$, they have peculiarities in 7 and 10 kOe for sample $B$. As we have mentioned above, the data for $\chi_1$, $\chi_2$ and $\chi_3$ are noisy and fluctuating at fields lower than 4 kOe at 4.5 K and 2 kOe at 5.5 K due to magnetic flux jumps. The behavior of the ac response in the SF mode has some similar features for both samples. Thus, an increase of the excitation amplitude and frequency leads to a decrease of $\chi_1^{''}$ in fields down to $H_{c2}$ and $\chi_2$ in the whole field range. The reason for this behavior is the following. The main physical parameter defining the difference between the PBP and SF modes is $Q=\frac{\dot{H}_0}{\omega h_{ac}}$ \cite{MAX,FINK2}. The PBP mode corresponds to $Q=0$. Parameter $Q$ decreases with the excitation amplitude and/or frequency tending to zero. This is why $\chi_1^{''}$ and $\chi_2$ decrease with $h_0$ and $\omega$ and
in consequence of this in the SF mode perturbation theory is not applicable. In the limiting case of high frequencies, for example, in the GHz range, a swept field with sweep rate of few tens or hundred oersted per second does not affect the ac response \cite{VAT}.

The ac response of sample $A$ in the SF mode is similar to that reported in our previous papers \cite{Genkin1,Katz1}. In this sample we observe a smooth field dependence of $\chi_1^{''}$, $\chi_2$ and $\chi_3$. The models proposed in \cite{Genkin1,Katz1} can explain the experimental data for sample $A$ in magnetic fields lower and higher than $H_{c2}$. The case with the sample $B$ is more complicated. It turned out that $\chi_1^{''}$ at magnetic fields of 4 kOe ($H_0< H_{c2}$) is lower for sample $B$ than for sample $A$, see Fig. \ref{f11a}$c$. The following may be the reason for this. Vortex pinning and the current induced by ac and swept fields play an important role in ac response in a swept magnetic field \cite{Genkin1}. The area under the row of antidots is much smaller than the total film area. This is why vortex pinning by this row antidots cannot explain loss reduction. At the same time the total induced current is lower in sample $B$ than in sample $A$, Fig. \ref{f3}. This reduces the forces dragging vortices into the substrate and leads to loss reduction \cite{Genkin1}. The jump at $H_0\approx 5$ kOe takes place only for sample $B$, see Figs. \ref{f9a} and \ref{f10a}. At fields higher than the jump field the losses for both samples at $h_{ac} =0.2$ Oe are equal, panels $b$ and $c$ of Fig. \ref{f11a}. The weakening of pinning in high magnetic fields could be a cause for such behavior.

The nature of the jump of $\chi_{3,2}$ in magnetic fields of 10 kOe (see panels \textit{b} of Figs.\ref{f9a} and \ref{f10a}) for sample $B$ in SF mode is not clear. ac amplitude is not smeared this jump completely in contrast with $\chi_1^{''}$, see panels \textit{a} and \textit{c} of Fig.\ref{f11a}. This jump takes place in magnetic fields near $H_{c2}$ of sample $B$.  Decreasing of ac losses and harmonics jump near $H_{c2}$ in a swept field was observed in single crystal Nb \cite{GENKIN22,MT}. However, single crystal Nb has a well defined vortex structure and $H_{c2}$ but it is not the case with our sample.

\section{conclusion}

We have studied the dc and ac magnetic properties of thin-walled cylinders of superconducting Nb with and without a row of antidots.  Experiment showed that the critical current density is higher in the isthmus between antidots than in the film itself. The dc magnetization curves demonstrate an "avalanche''-like penetration of the magnetic flux into the cylinder for both samples. The effect was observed at a temperature of 4.5 K and completely disappeared at 7 and 5.5 K for samples $A$ and $B$, respectively.
Such a behavior resembles a thermomagnetic instability of vortices but it was observed in fields below $H_{c1}$ of the films, i.e. in a vortex-free state. The effect of end faces, consisting in that the magnetic force lines is bending near the sample ends, could be another reason for flux jumps. The influence of the sample end faces on the flux jumps in such samples has to be studied using a local probe technique.

The ac response of thin-walled cylinders with and without antidots is strongly nonlinear and perturbation theory cannot explain the experimental data. The ac response of $A$ and $B$ samples is similar in the point-by-point mode. However, in the swept field mode there is a qualitative difference between losses for samples $A$ and $B$. Thus, at low magnetic fields, losses in sample $B$ are lower than in sample $A$. There are jumps in $\chi_1$, $\chi_2$ and $\chi_3$ in high magnetic fields for sample $B$, but these quantities are smooth functions of the magnetic field in sample $A$.

We demonstrate that field dependences of $\chi_1^{''}$ and $\chi_3$ have the same shapes in the point-by-point mode with high accuracy. In the swept field mode the shapes of  $\chi_1^{''}$ and $\chi_3$ are the same in the vicinity of $H_{c3}$.  This similarity has yet not been proved in the case of strong nonlinear response that we encounter in our experiment.

The models developed in \cite{Genkin1,Katz1} could describe the ac response of the as-prepared sample. However, these models are not applicable to the sample with a row of antidots. New models for samples with antidots have to be elaborated. As well as further experimental studies of samples with different lengths, wall thicknesses, sizes and geometry of antidots row or array have to be carried out.

\section{acknowledgments}
We thank J. Kolacek, P. Lipavsky and V.A. Tulin for fruitful discussions.
This work was done within the framework of the NanoSC-COST Action MP1201.
Financial support of the grant agency VEGA in projects nos. 2/0173/13 and 2/0120/14 are kindly appreciated.

\end{document}